\documentclass{aa}  
\usepackage{graphicx,natbib,ulem}
\usepackage{txfonts}
\usepackage{color}
\begin{document}
  \titlerunning{The circumbinary disk of SS 433}
\authorrunning{ Bowler}
   \title{More on the circumbinary disk of SS 433 }

   \subtitle{}

   \author{M. G.\ Bowler \inst{}}

   \offprints{M. G. Bowler \\   \email{m.bowler1@physics.ox.ac.uk}}
   \institute{University of Oxford, Department of Physics, Keble Road,
              Oxford, OX1 3RH, UK}
   \date{Received 21 December 2010 / Accepted 30 May 2011}

 
  \abstract
   {Certain  lines in spectra of the Galactic microquasar SS\,433, in particular  the brilliant Balmer H$\alpha$ line, have been interpreted as emission from a circumbinary disk. In this interpretation the orbital speed of the glowing material is in excess of 200 km s$^{-1}$ and the mass of the binary system in excess of 40$M_\odot$. A very simple model for excitation of disk material is in remarkable agreement with the observations, yet it seems that the very existence of a circumbinary disk is regarded as controversial.} 
  {To investigate whether  analysis of optical data from H$\alpha$ and He I spectral lines in terms of a  model, in which the disk is excited by radiation from the close environment of the compact object, can further illuminate the origin of these split spectral lines.}
   { A model in which the excitation of any given patch of putative circumbinary disk material is proportional to the inverse square of its instantaneous distance from the compact object was constructed and compared with published spectra, taken almost
     nightly over two orbital periods of the binary system. The H$\alpha$  and He I lines were analysed as superpositions of Gaussian components.  }  
  { The new model provides an excellent description of the observations. The variations of the H$\alpha$ and He I spectra with orbital phase are described quantitatively, provided the radius of the orbit of the emitting ring is not much greater than the radius of the closest stable circumbinary orbit.  The observed variations with orbital phase are not consistent with an origin in a radially expanding ring.}
{ The new analysis has greatly strengthened the case for a circumbinary disk orbiting the SS 433 system with a speed of over 200 km s$^{-1}$ and presents supposed alternative explanations with major difficulties. If the circumbinary disk scenario is essentially correct, the mass of the binary system must exceed 40 $M_\odot$ and the compact object must be a rather massive stellar black hole. This possibility should be taken seriously.}


   \maketitle
%

\section{Introduction}

 This paper is concerned with the origin of split lines in the optical spectrum of the Galactic microquasar SS 433, which have been attributed to a circumbinary disk. It contains a development of the model presented in Bowler (2010a), to which the reader is referred for the story so far and numerous relevant references. The H$\alpha$ and He I spectra here addressed are very usefully displayed in Schmidtobreick \& Blundell (2006), Fig.2.

      SS 433 is very luminous and unique in its continual ejection
of plasma in two opposite jets at approximately one quarter the speed
of light. The system is a 13 day binary and probably powered by supercritical accretion on the compact member from its companion. The orbital speed of the compact object is fairly well established but in order to determine its mass either a measurement of the orbital velocity of the companion is needed or a measure of the total mass of the system. 

   Stationary emission lines in the spectra of SS 433 display a persistent two horn structure of just the kind expected for emission from an orbiting ring, or a disk, seen more or less edge on. The horn separation corresponds to a rotation speed in excess of 200 km s$^{-1}$ and attributed to material orbiting the centre of mass of the binary system implies a system mass in excess of 40 $M_\odot$. The H$\alpha$ spectra were originally discussed in  Blundell, Bowler \& Schmidtobreick (2008). Departures from the pattern expected for a uniformly radiating ring are present and are more pronounced in He I emission lines. These departures were explained in terms of emission being stimulated by some kind of spotlight rotating with the binary, rather like the beam from a lighthouse (Bowler 2010a). That paper presented an analysis in terms of a very simple spotlight model; too simple to be realistic and yet  yielding an astonishingly good description of the observations. Thus these observations fix rather well the mass of the SS 433 system and hence of the compact object, subject only to the proviso that the two horned structure in H$\alpha$ and He I is indeed produced in an orbiting circumbinary ring, perhaps the inner rim of a circumbinary disk. This notion has been greeted with some scepticism, one correspondent going so far as to describe it as "very speculative" and suggest a radially expanding ring as a more plausible explanation for the two horned structures (D. R. Gies, personal communication).
   
       My purpose here is to construct a realistic model of the appearance of spectral lines emitted from a disk excited by intense radiation from the environs of the compact object and compare these features both with the data and with such alternative mechanisms as have been suggested (although, so far as I am aware, no other suggested mechanism has been properly worked out).  The model describes the data very well and its development has made it clear that the data do not agree with a radially expanding ring.

    \section{The model and the data}
    
     If a circumbinary ring in a circular orbit about the system is viewed close to edge on, emission  lines exhibit a two horn structure. If the ring is emitting with intensity independent of azimuth, the relative intensities of the horns stay equal as a function of time and the separation of the horns gives the orbital speed of the ring. The essential description of the spectral distribution produced by a circumbinary ring in a circular orbit, close to edge on, is found in Eqs. 1 and 2 of Bowler (2010a). In contrast, were the circumbinary ring being excited by a narrow lighthouse beam rotating with the period of the system, the resulting spectrum would vary with time just as emission of the same spectral line from a single body orbiting the centre of mass of the binary, with the speed of the circumbinary material and a 13 day period. The appearances of observed spectra are between these two extremes; H$\alpha$ close to uniform emission and He I distorting with time from red to blue and back with a period of 13 days. The reader is recommended to inspect Fig.2 of Schmidtobreick \& Blundell (2006); that figure is vivid and most informative. The two horns in H$\alpha$ are separated by roughly 7 \AA\ and their intensities vary by about 0.2 of their amplitudes, in antiphase and with a 13 day period. The He I spectra show much greater variation of the red and blue horns over the period of the binary orbit. The separation between the wavelengths of the red and blue horns, when each is dominant, is again roughly 7 \AA\ but the intensities vary at these wavelengths by a factor of about 3. Thus the excitation of the two horned structures varies with a period of 13 days and if the excitation mechanism is the same for H$\alpha$ and He I then the H$\alpha$ source has the longer memory. These are the qualitative features which any model for the two horned structures must explain quantitatively. My original simple model (Bowler 2010a) assumed excitation of an orbiting circumbinary ring by a narrow spotlight rotating with the binary, but  emission from excited material decaying exponentially with time. The data were well represented by decay times of about 4 days for He I and more like 14 days for H$\alpha$. Thus emission as a function of azimuthal angle was represented by a sharp leading edge followed by exponential decay of the excited material, the whole rotating with time. This was a simple and flexible parametrisation of the sort of effect likely to result and did not embody any specific assumptions about the machinery producing the hot spot, nor of the radius of the ring. The sharp leading edge as a lighthouse beam passes (see Fig.3 of Bowler 2010a) is scarcely realistic, yet the H$\alpha$ spectra are best described if the leading edge is no more than a day ahead of the compact object and 2 days ahead for He I. This suggests that at any moment the most intense radiation may be coming from that part of the putative circumbinary disk closest to the compact object.
    
    \begin{figure}[htbp]
\begin{center}
{ \includegraphics[width=9cm,trim=0 0 0 2]{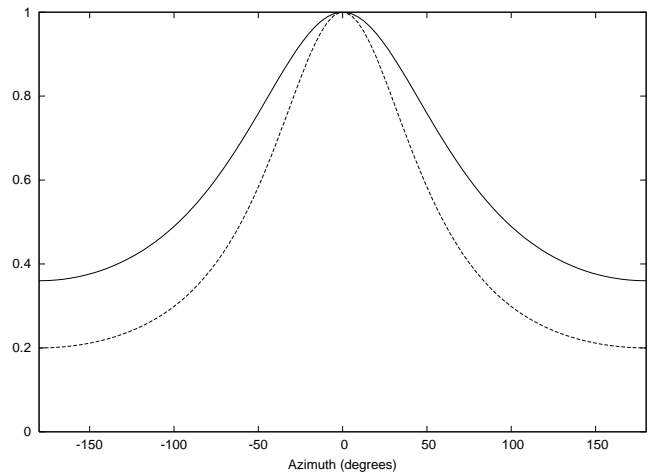} }        
   \caption{The figure shows the illumination of the circumbinary disk by an effectively point source located at the compact object. The azimuthal angle is measured from the line connecting the companion with the compact object; thus intensity 1 at azimuthal angle 0 degrees corresponds to illumination of that patch of the circumbinary ring closest to the compact object. The illumination pattern rotates round the ring with a period of 13.08 days. The wider distribution corresponds to a geometry for equal masses with the orbital radius of the compact object $A/2$ and of the circumbinary ring $2A$. The narrower corresponds to a radius of the ring orbit $1.6A$ and a somewhat smaller mass ratio. Both geometries are plausible.
   }
\label{fig:ideagram}
\end{center}
\end{figure}

     I have therefore assumed in this work that at any moment the circumbinary ring is illuminated (and thereby excited) proportional to the inverse square of the distance between the circumbinary ring material and the compact object (incident angle has little effect for plausible geometries). Thus the original single hotspot of negligible dimensions has been replaced by a distribution; the model illumination as a function of  azimuthal angle relative to the compact object is displayed in Fig.1 for two sample geometries. The distribution of the radiation source round the ring in the old model has been convolved with these inverse square excitations of the ring. The broader distribution in Fig.1 is for a geometry in which the circumbinary ring orbits at a radius of $2A$ and the compact object at a radius of $A/2$ about the centre of mass of the binary, where $A$ is the binary separation; the ratio $q$ of the mass of the compact object to that of the companion is 1. The distance between points on the ring and the compact object thus varies between $3A/2$ and $5A/2$ and the illumination between 1 for the nearest point and 0.36 for the furthest. The narrower distribution is for a geometry in which the emitting material is orbiting at $1.6A$ (rather within the radius of the last stable orbit) and $q$ is rather smaller than 1. I regard any greater variation in illumination due to inverse square effects alone as rather implausible. In implementing this function I also cut out the region within $30^{\circ}$ of the line joining the compact object to the companion, as a crude representation of eclipse effects lasting about 2 days.
   
    \begin{figure}[htbp]
\begin{center}
   \includegraphics[width=6cm]{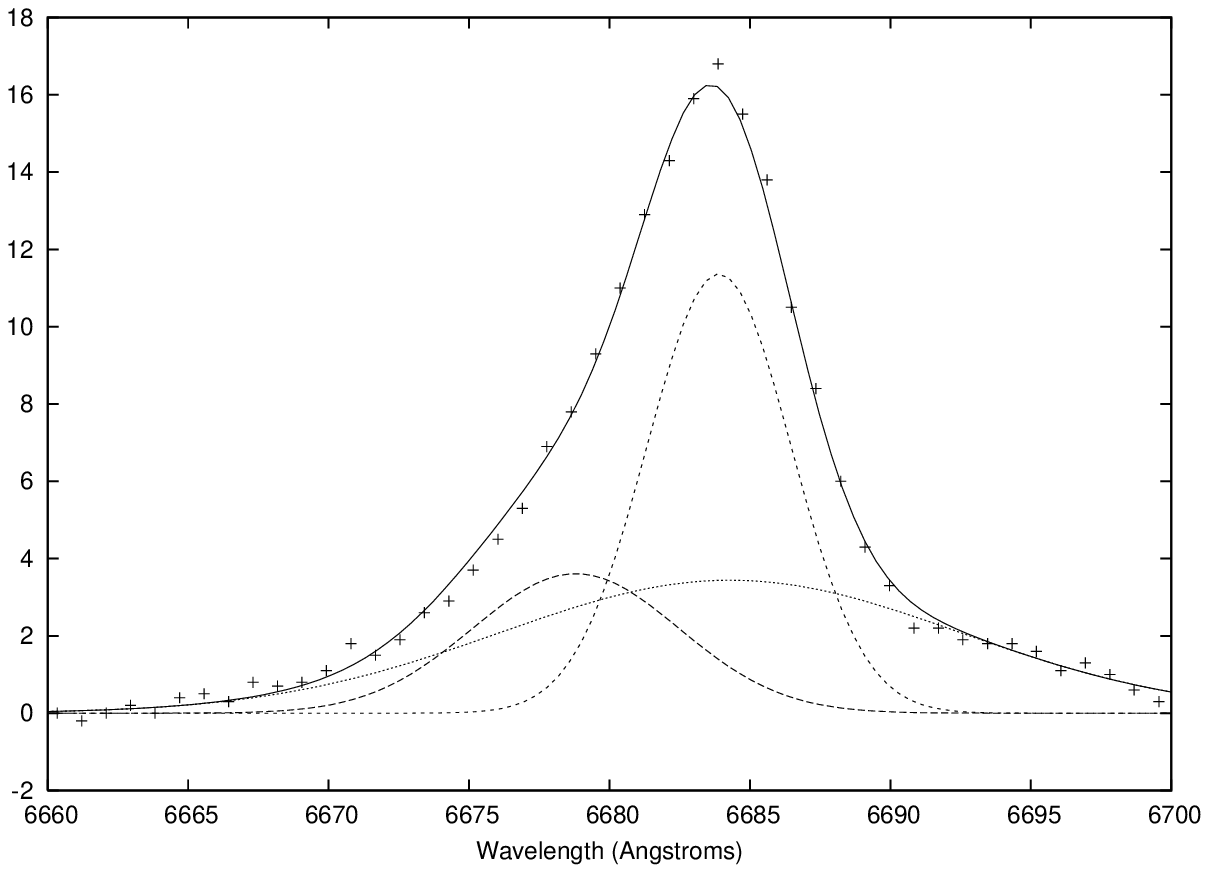} 
   \includegraphics[width=6cm]{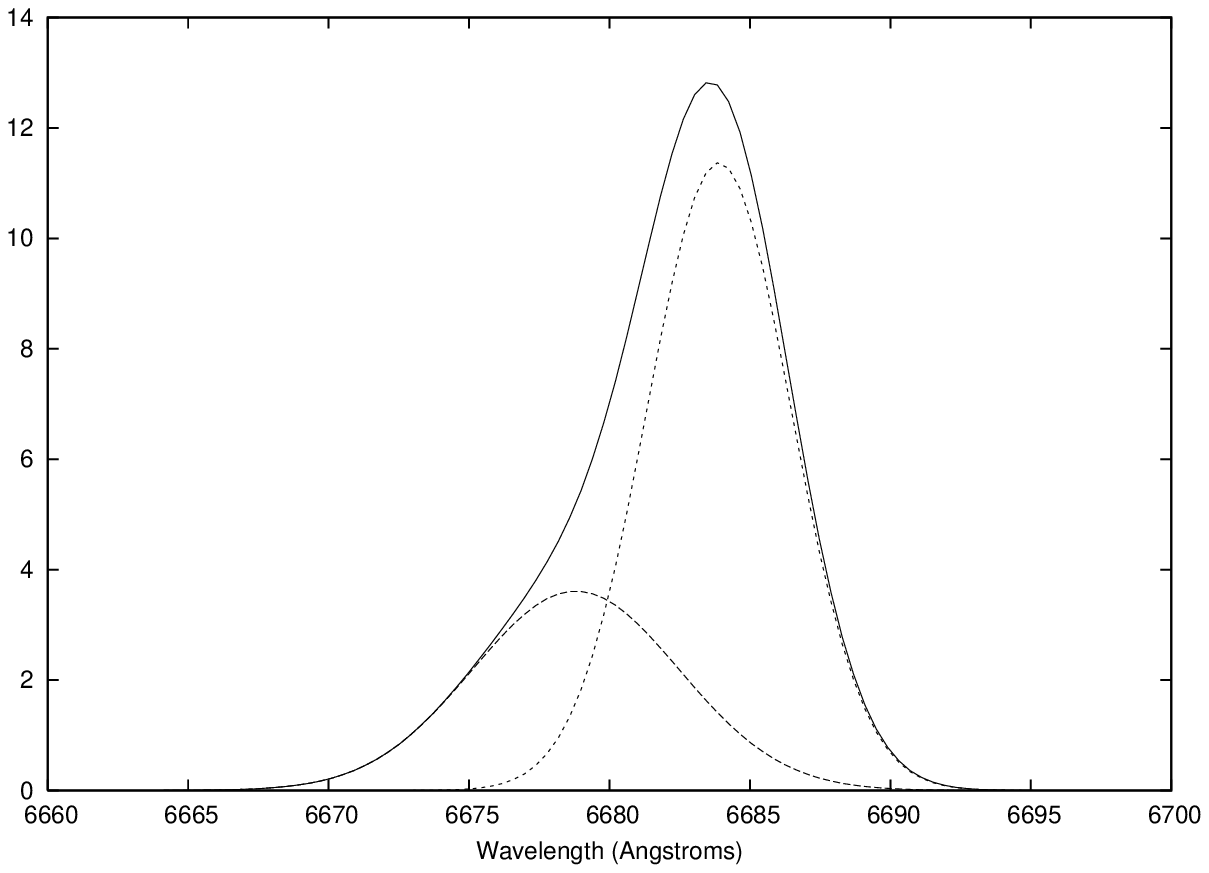}
   \includegraphics[width=6cm]{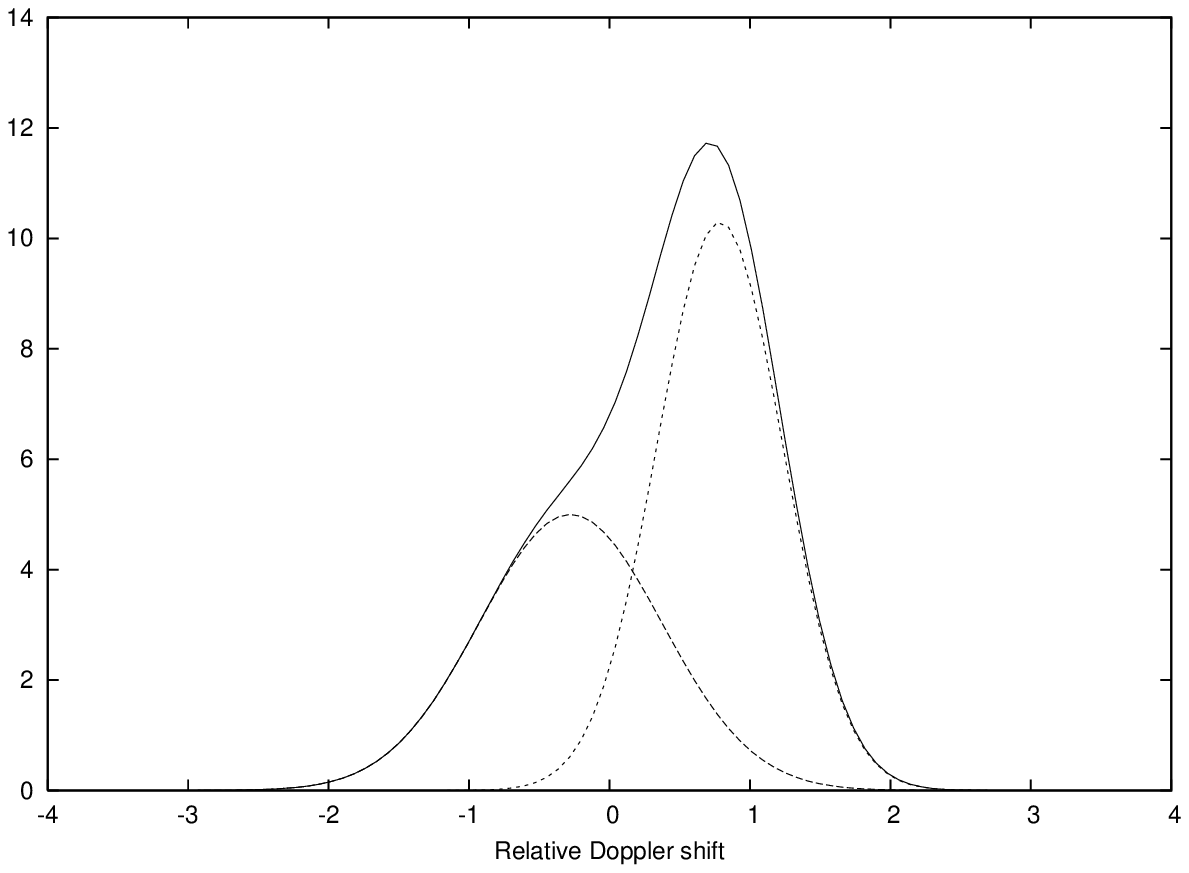}
\caption{ This figure illustrates the quality of the He data and of its decomposition in terms of Gaussian components. The upper panel displays the digitised He I 6678 \AA\ spectrum for JD +253.5, close to the extreme red shift exhibited, together with the three components of the fit and the resulting envelope. The middle panel shows the two narrow components and their resultant, the broadest Gaussian having been stripped out. The lowest panel is the result of analysing the spectrum from the model described in the text, for approximately the same extreme. The system is about 1.4 days before orbital phase zero. No attempt has been made to fit the model in detail but the similarity between the modelled spectrum and the reduced data shown in the middle panel is evident. A comparison of the various panels shows that the relative Doppler shift for $x=1$ corresponds to red shift of above 200 km s$^{-1}$.
 }
\label{fig:disc}
\end{center}
\end{figure}

    \begin{figure}[htbp]
\begin{center}
 \includegraphics[width=6cm]{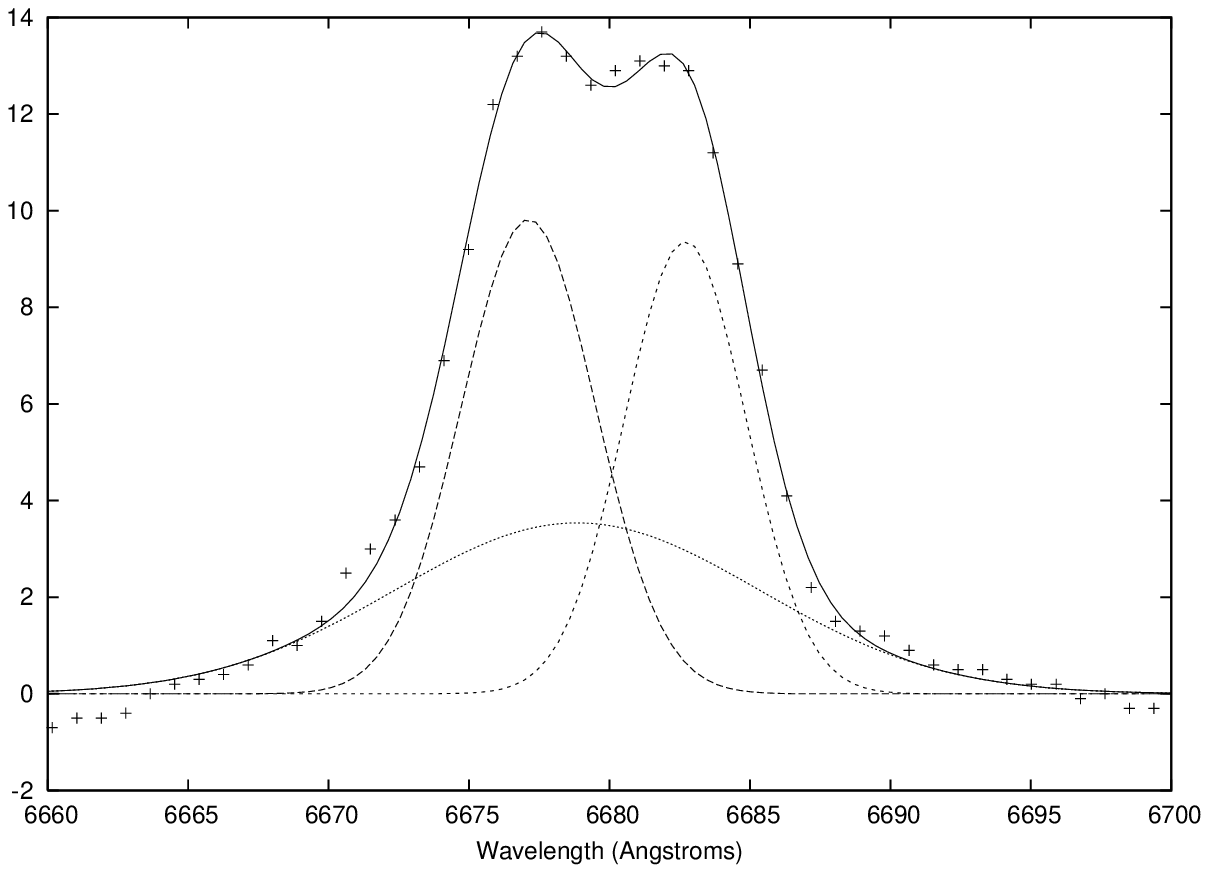}
  \includegraphics[width=6cm]{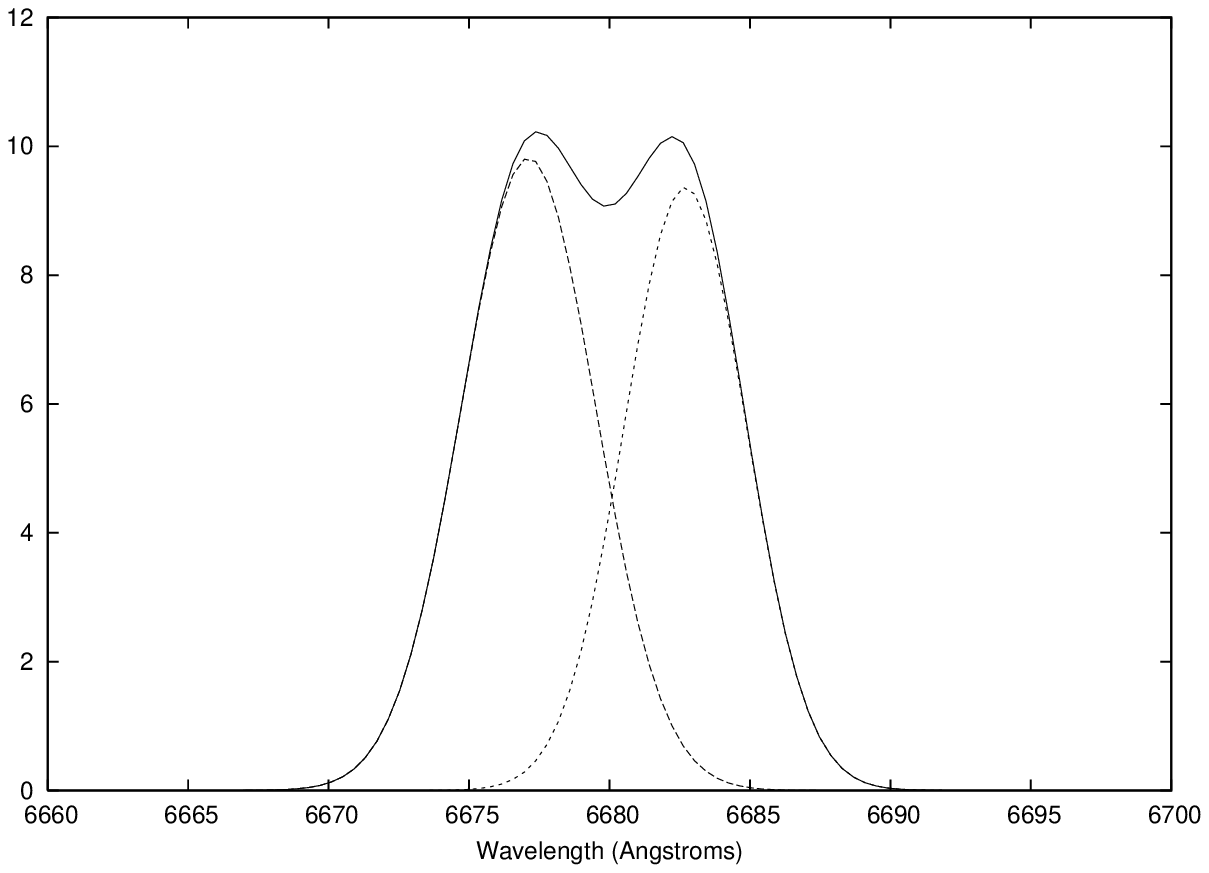}
   \includegraphics[width=6cm]{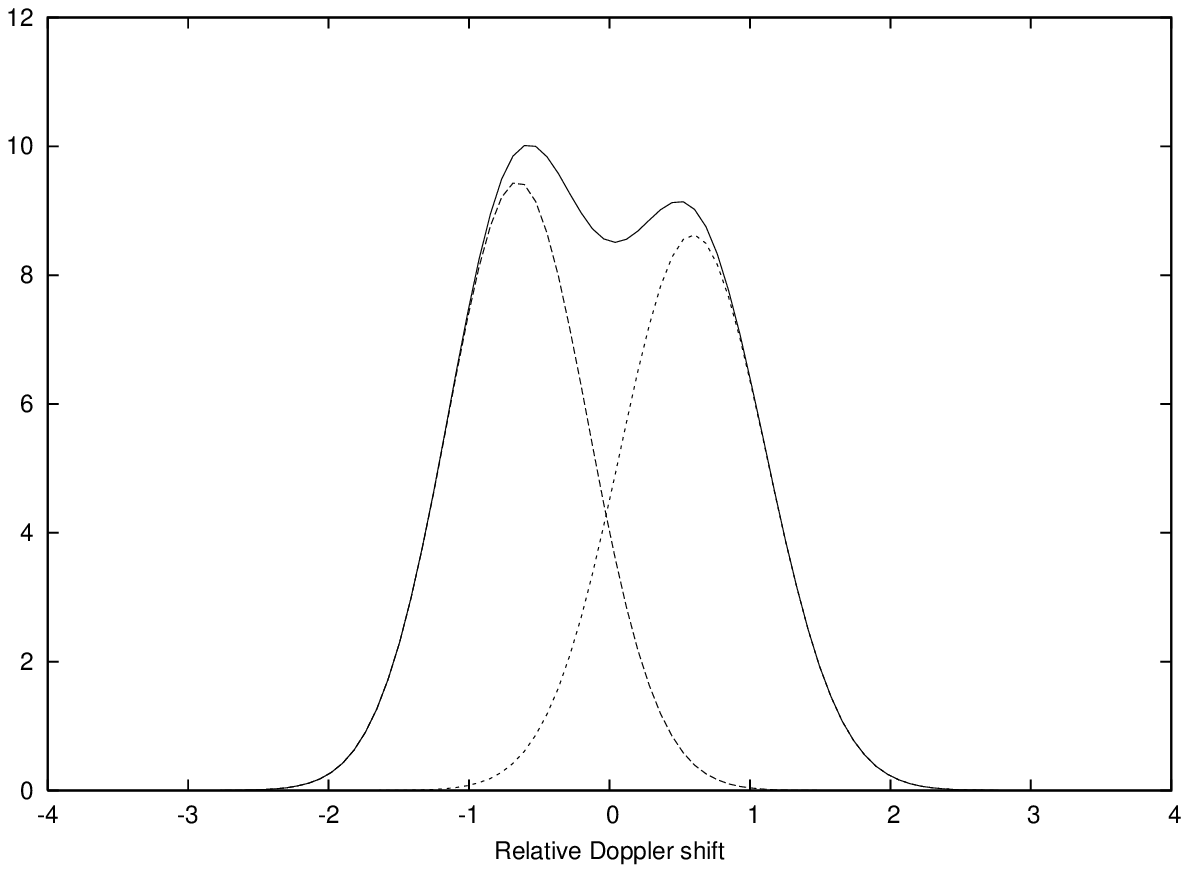}
\caption{ This figure is a companion to Fig. 2 but the data are for JD +248.5, just after orbital phase 0.5. In Fig. 2 the red horn dominates the He I 6678 \AA\ spectrum; here the red and blue horns are of approximately equal intensity in both the data and the model.}
\label{fig:248}
\end{center}
\end{figure}

 \begin{figure}[htbp]
\begin{center}
   \includegraphics[width=15cm,trim=0 0 0 2]{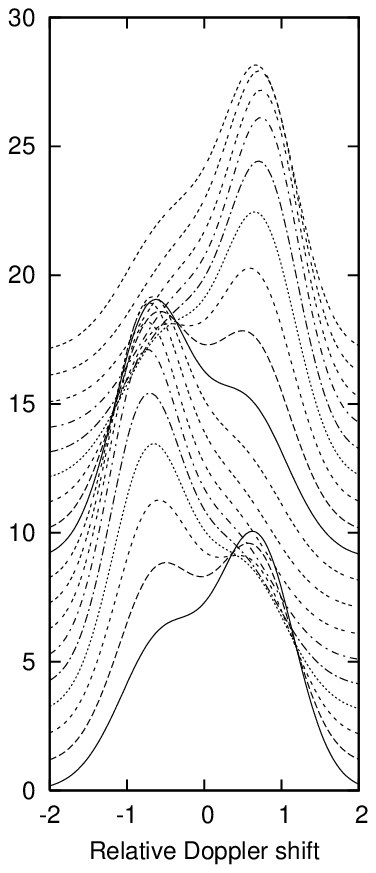} 
   \caption{ Model spectra for He I, generated every 20$^{\circ}$ of orbital phase. The lowest spectrum is for orbital phase 0; successive spectra are successively displaced upwards by 1. The top spectrum is for 340$^{\circ}$ and the model spectrum shown in Fig. 2 immediately precedes it. The model spectrum shown in Fig. 3 is also included. This compilation can be compared with the observed spectra in Fig. 2 of Blundell \& Schmidtobreick (2006).}
\label{fig:timesequence}
\end{center}
\end{figure}

      As in Bowler (2010a), I generated the spectral shape for orientations relative to the line of sight every $20^{\circ}$, as a function of the variable $x$, the red shift as a fraction of the maximum possible. Thus a value of $x=-1$ corresponds to radiation from material in an orbiting ring approaching tangential to the line of sight. The generated spectra were then decomposed into a superposition of two Gaussian functions; examples are shown in the lowest panels of Figs. 2 \& 3. Fig. 4 displays a compilation of the 18 model spectra calculated for He I.
      
      Because of the distributed excitation function, the rapid variations in the Paper I patterns are smoothed and it is necessary to introduce faster exponential decay of the source intensities  in order to represent the spectra. Because the intention is to attempt a more accurate description of the data, I also refitted all the digitised He I spectra. These are noisier than H$\alpha$ and have a narrower broad component. In some cases I found fits not only formally better but also more acceptable in that the broad component was not abnormally large in amplitude. In this paper I have used the new fits to He I 6678 \AA\ and for the model the narrower excitation function in Fig.1.
      In Figs. 2 \& 3 I show examples of fits to the data in terms of Gaussian components and  comparisons with model predictions. Fig. 2 is for an extreme shift of the He I line to the red (JD 2 453 000 +253.5) and for the corresponding configuration in the model; Fig. 3 is for roughly equal heights of the two horns (JD +248.5). In both cases the model describes the contribution of the two narrower components in the data rather well. The assumed exponential decay time for He I is about 1 day and so the emission function is dominated by the shape of the excitation function. The detailed comparisons of data with model shown in Figs. 2 \& 3 are placed in context by the sequence of model spectra over a full orbit, Fig. 4. This figure has been constructed to be in the same format as the compilations of data shown in Fig. 2 of Blundell \& Schmidtobreick (2006). It is quite obvious from such a comparison that the excellent agreement detailed in Figs. 2 \& 3 is maintained over the full orbit.

 \section{Results of matching the new model to the data}
     In the upper panel of Fig. 5 I show the Doppler speeds of the narrow Gaussian components fitted to the He I 6678 \AA\ data, day by day for JD +245 to +274. This panel may be compared with the lower panel in Fig. 1 of Bowler (2010a). There are differences, but mostly within the uncertainties specified in that paper. The lower panel is calculated under the assumption that the model for excitation by irradiation from the vicinity of the compact object is correct, with a short decay time. The speed of orbital rotation attributed to the ring is 230 km s$^{-1}$ and an assumed systemic speed of 70 km s$^{-1}$ has been added.  This lower panel may be compared with the original model calculation;  lower panel in Fig. 2 of Bowler (2010a). 
     
     The new model captures the gross features of the data rather successfully, in particular the snaking oscillation in both the red and blue components. It is important to note that the phase of these oscillations is not an adjustable parameter and the agreement with the data is evidence that the hottest spot on the circumbinary ring rotates with the phase of the compact object. The hottest spot is given by the line between the companion and the compact object, extrapolated to the ring, and lies on the line of sight when the compact object and disk eclipse the companion (orbital phase 0.5); JD + 248.36  according to the Goranskii ephemeris (Goranskii et al 1998).
     
      The comparison may also be carried out in terms of the mean of the Doppler shifts and, separately, half the difference. For a uniformly glowing ring the former would be the systemic velocity and the latter the rotation speed of the orbiting ring. Fig. 6 shows the variation of the apparent rotational speed with time and Fig. 7 the apparent systemic recession. These figures may be compared with the corresponding Figs.5 and 6 in Bowler (2010a). As in Fig. 5, the model calculations are made for the narrower source distribution from Fig.1 and a short decay time of about 1 day. The assumed rotational speed of the ring was 230 km s$^{-1}$, which gives the best representation of the He I data, as for the model in Bowler (2010a).  The agreement between the model and the data is very good; in particular the phasing of the variations relative to Julian date is not a free parameter, yet here the model and the data agree to within a day.
 
\begin{figure}[htbp]
\begin{center}
   \includegraphics[width=0.49\columnwidth,trim=0 0 0 2]{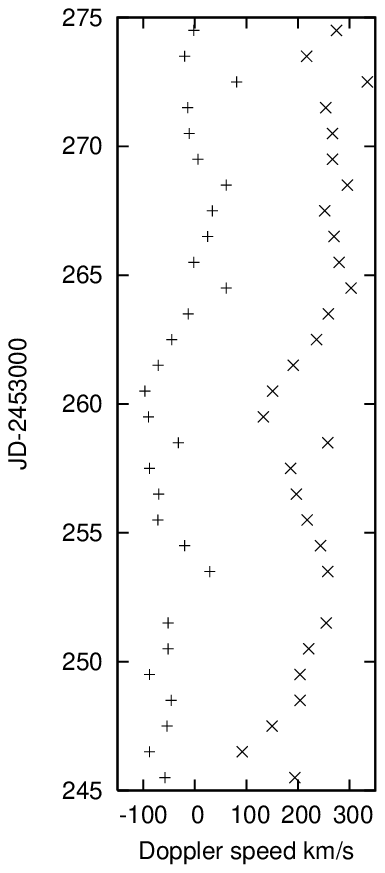} 
   \includegraphics[width=0.49\columnwidth,trim=0 0 0 2]{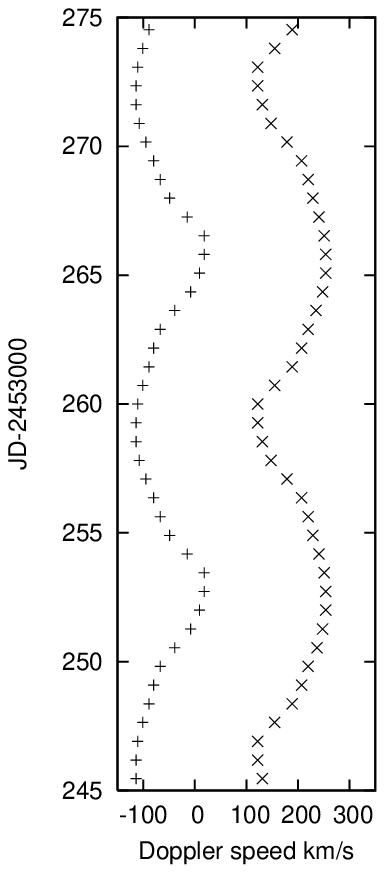}
   \caption{ Doppler speeds of the fitted blue and red Gaussian components attributed to the circumbinary disk are shown in the left panel for He I data and in the right for the model described in the text. Julian date increases vertically and in both panels the bluer component is denoted by + and the redder by x. The model reproduces the gross features of the data rather well, in particular the snaking from blue to red and back.}
\label{fig:timesequence}
\end{center}
\end{figure}

It is also important that with such a short decay time there is no freedom in the model to accentuate the magnitude of the variations of the rotational and recessional velocities with time, because these details are dominated by the geometry of the source and not by the decay time. If the wider source distribution in Fig.1 were employed, the depth of the minima in rotational speed would decrease by 10 km s$^{-1}$ and the amplitude of  the recessional oscillations by 25 km s$^{-1}$. Obviously, as the source distribution flattens out completely the oscillations in both quantities vanish. Thus for geometries giving  flatter source distributions than those illustrated in Fig.1 the agreement between model and data would become progressively worse for He I.  An assumed decay time greater than 1 day would not succeed in producing oscillations of the magnitudes shown in Figs. 5-7. The new model also matches the H$\alpha$ data, originally presented in Blundell, Bowler \& Schmidtobreick (2008) and again in Bowler (2010a). The new model has less flexibility than the old, but a longer damping time is again needed to generate a sequence of spectra with comparatively small oscillations in the heights of the H$\alpha$ horns. This is reflected in very small snaking in the H$\alpha$ analogue of Fig. 5; to a first approximation the red and blue components attributed to the circumbinary disk run railroad straight over 30 consecutive days (and can be traced beyond JD +294 until flaring leads to some obscuration, Bowler 2010b). The H$\alpha$ data are best represented with a decay time of about 3 days and as in Bowler (2010a) a rotational speed rather over 250 km s$^{-1}$. As the source distribution is flattened the indentations in rotational speed drop in magnitude faster than those in recessional speed and as for He I the narrower distribution in Fig. 1 is preferred over the broader.  As for the He I data, the phase of the indentations modelled assuming that the hottest spot on the circumbinary ring is that nearest to the compact object agrees with the data to within a day.
  
      It should be noted that because of the asymmetry in the emission functions which result from the decay of emission from each freshly stimulated region, the extreme velocities of the red and blue horns (see Fig. 3 for the redhorn) are not observed when the hottest spot on the circumbinary ring corresponds to material  approaching or receding fastest (orbital phases 0.75 and 0.25 respectively) but rather about 1.5 days later for He I and perhaps 2 days later for H$\alpha$.
    
\begin{figure}[htbp]
\begin{center}
   \includegraphics[width=9cm,trim=0 0 0 140]{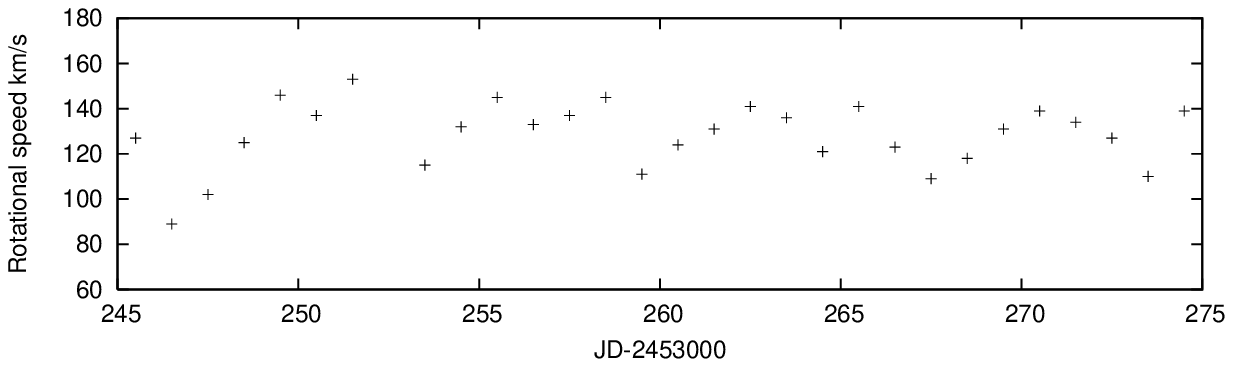}
   \includegraphics[width=9cm,trim=0 0 0 140]{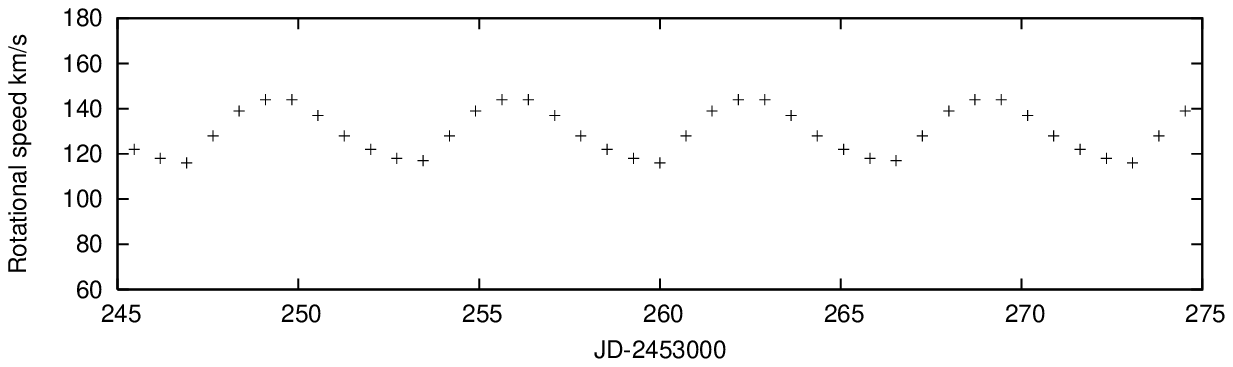}
   \caption{ The nominal rotational velocity of the circumbinary disk, as obtained from the differences between the red and blue components. The top panel is for He I 6678 \AA\ and the bottom panel is the model calculation. If the broader version of the model source is used, the depth of the rotational speed minimum is reduced by 10 km s$^{-1}$.}
\label{fig:timesequence}
\end{center}
\end{figure}

\begin{figure}[htbp]
\begin{center}
   \includegraphics[width=9cm,trim=0 0 0 120]{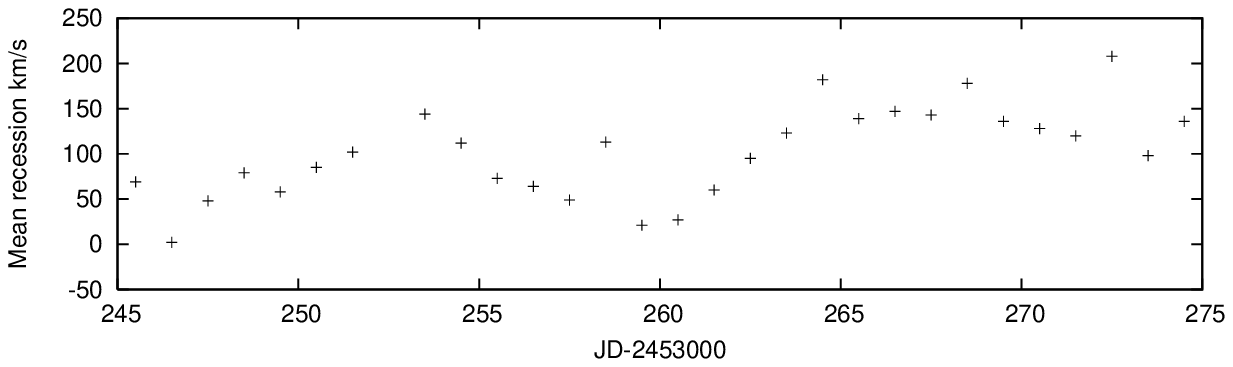}
   \includegraphics[width=9cm,trim=0 0 0 120]{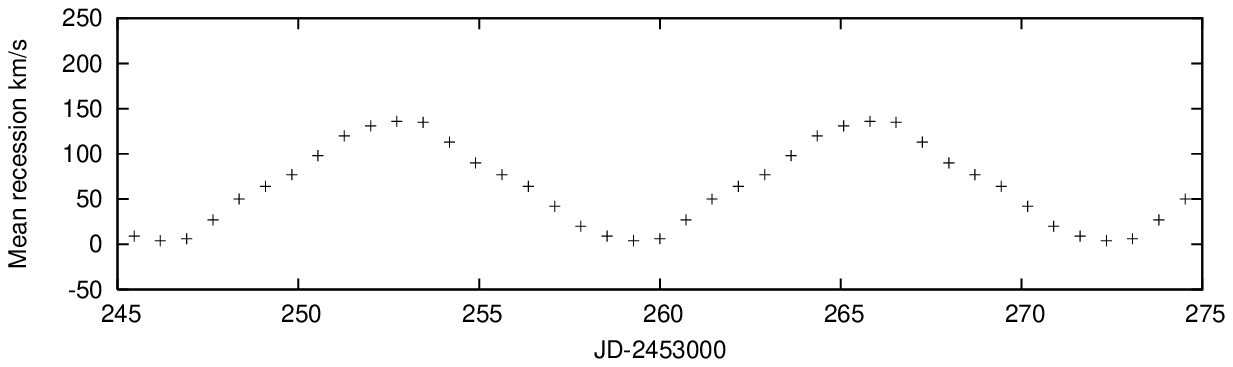}
   \caption{ The mean recessional velocity of the red and blue circumbinary disk components as a function of time. The top panel is for He I 6678 \AA\ and the bottom panel is the model calculation. If the broader version of the model source is used, the mean recession amplitude is reduced by 25 km s$^{-1}$.}
\label{fig:timesequence}
\end{center}
\end{figure}

  In summary, the new model in which radiation from the circumbinary ring is stimulated by irradiation from an effectively point source at the compact object is in very good agreement with the observed spectra. The geometry necessary for this agreement matches that inferred already (Blundell, Bowler \& Schmidtobreick 2008 and Bowler 2010a) and the phasing of the model to the motion of the compact object is correct. In Bowler (2010a) the lighthouse beam for He I was about 2 days in advance of the compact object but that for H$\alpha$ very close; these discrepancies have gone away in the new model presented here.

  \section{Discussion}
  
   The salient features of the data are that both H$\alpha$ and He I stationary spectral lines exhibit a two horned structure with about the same separation between the horns. The red and blue horn intensities oscillate in antiphase and the amplitude of oscillation is small for H$\alpha$; the horns in He I vary in intensity over an orbit by a factor of about 3. The differences can be explained in terms of freshly excited He I regions decaying with time (perhaps cooling) faster than H$\alpha$ regions.
  
  \subsection{ Strengths of the model}  
  
    These data have been interpreted in terms of a circumbinary ring; the H$\alpha$ observations in Blundell, Bowler \& Schmidtobreick (2008) and He I spectra in Bowler (2010a). The model presented in Bowler (2010a) was used to explain quantitatively the He I spectra and also features of H$\alpha$. In this paper the unrealistic model of a rotating searchlight beam exciting the material in an orbiting ring has been replaced by assuming that intense radiation from the vicinity of the compact object excites material in the ring, most strongly when closest. This model is still simple but may be more realistic. Both the original model and the new version can generate spectral shapes matching the sequences in both H$\alpha$ and in He I, qualitatively and quantitatively. The spotlight model required slightly different phases for H$\alpha$ and He I, in both cases a little ahead of the compact object.

  Thus the new model postulates that radiation from the circumbinary disk is dominated by excitation caused by radiation from the vicinity of the compact object. Because that irradiation is spread right round the circumbinary ring and because a uniform source will not generate the sort of structure shown in Figs. 5-7, it was not clear in advance of detailed calculations that this postulate would actually work. These calculations have in fact shown that for a circumbinary ring orbiting at a radius of about $2A$ this model is viable, but a substantially larger ring radius would not work, given the known mass function and the orbital speed of the circumbinary material. Both the H$\alpha$ and He I spectra have oscillations correctly phased to the orbital motion of the compact object. Thus a realistic and constrained model has provided a more satisfactory description of those aspects of the H$\alpha$ and He I spectra attributed to a glowing circumbinary disk. The very great stability exhibited by the H$\alpha$ components (Blundell, Bowler \& Schmidtobreick 2008, Bowler 2010a; there appears to be evidence for stability over a much longer period in Fig.2 of Li \& Yan 2010) in itself  is a compelling argument for a circumbinary disk origin; the new model locks better together the H$\alpha$ and He I components, thereby increasing the probability that they share the same circumbinary origin. Thus the case for a circumbinary disk orbiting at approximately $2A$ (or rather less) and excited by radiation is strengthened.
  
  \subsection{ Possible difficulties}
  
   Both the old and new models are possibly unsatisfactory in two ways. The first is the introduction of characteristic cooling times for the H$\alpha$ and He I sources, parameters  tuned to match the data and which are different by a factor of about 3 in both models; this is not explained. Perhaps it is H$\alpha$ that is exceptional - the Paschen lines also have a two horned structure and shimmy like He I (Blundell, Bowler \& Schmidtobreick 2008). The second feature is that within the models H$\alpha$ is best represented by an orbital speed of rather above 250 km s$^{-1}$ and He I by a speed rather below 250 km s$^{-1}$. These two aspects might be related, either through inadequacies in the modelling or through the H$\alpha$ and He I lines being formed in slightly different regions under slightly different physical conditions. However it may be, the emission function for H$\alpha$ has to be rather less localised in azimuth than for He I. If this smoothing is not a manifestation of a longer decay time it would have to be due to another effect, perhaps a degree of saturation of H$\alpha$ emission or, even less plausibly, H$\alpha$ emission lines being formed in a region rather further out than $2A$.
  
  Absorption spectra of SS 433 show features characteristic of a mid A-type star and these absorption lines, observed over more than a quarter of an orbit, exhibit a 13 day period with an amplitude of 58 km s$^{-1}$ (Hillwig \& Gies 2008, Kubota et al 2010). If these absorption lines are formed in the atmosphere of the companion, then the origin of the two horned structures which are the subject of this paper cannot possibly be in a circumbinary disk. However, the kinematics are such that absorption of continuum radiation from the companion in the orbiting circumbinary material would look exactly like a companion orbiting at $\sim$ 60 km s$^{-1}$; this possibility is discussed in more detail in Bowler (2010a, 2010c).
 
 \subsection{On the question of wholly different explanations}
  
 The success of the new model (which cannot be expected to be right in every detail) has emphasised the case against supposedly alternative explanations. I am aware of two classes of such alternatives. The first is the possibility that the structures attributed to a circumbinary disk might have an origin instead in gases ejected through the L2 point in the binary system and heading outwards, yet still sufficiently close to the binary for the loss of kinetic energy and curvature of a parabolic or hyperbolic orbit to be apparent. This I considered in Bowler (2010a); my calculations indicated that such an origin would not be consistent with the extreme stability of the H$\alpha$ railroad but might have to be considered for He I. The new model links even better H$\alpha$ and He I, thereby further decreasing the plausibility of L2 ejecta being the immediate source, as opposed to being stimulated after taking up residence in a circumbinary disk.
     
      The second class of possible origins is that rather than being formed in an orbiting circumbinary ring or disk, the features so attributed might be formed in material moving radially outward from the binary system. Here there are three categories. (i) For a ring of material expanding outward and excited uniformly in azimuth, the structure of a radiated spectral line is indistinguishable from that of a circumbinary ring orbiting with that expansion speed; I had dismissed the notion of an expanding ring fairly close in, on the grounds of continuity and stability of the H$\alpha$ railroad (Bowler 2010a). (ii) If wind blows out with radial symmetry perhaps the spectral lines are formed at a radius where there is always wind moving at about 250 km s$^{-1}$? This might give long term stability. D. R. Gies (personal communication) has suggested that wind from the disk might contain one or more ringlike enhancements launched from the inner (accretion) disk - this might fail the stability test. Formation of these two horned structures in the disk wind seems an unlikely origin for the following reasons. Underlying the two horned structures are much broader components formed in the wind from the disk;  they retain a substantial memory of the orbital motion of the compact object and its disk, which the two horned structures do not share (Blundell, Bowler \& Schmidtobreick 2008). Finally, Perez \& Blundell (2010) have used observed Balmer decrements to argue that the origin of the two horned structure in H$\alpha$ lies beyond the wind. (iii) There is certainly material expanding outwards from the binary and at considerable distances the motion of such material becomes essentially purely radial. If the emission lines of H$\alpha$ are formed at a suitable large distance stability is possible because the material is continually refreshed (see Fabrika 1993).
     
  \subsection{ Radial expansion excluded}
  
   The new analysis presented in this paper is strongly against an origin of the two horned spectral structures through any radial expansion mechanism, such as the three mentioned above. 
      A glowing ring orbiting at 250 km s$^{-1}$ is indistinguishable from a glowing ring expanding radially at 250 km s$^{-1}$ - provided that in both cases the glow is uniform in azimuth around the ring. The phenomena here discussed do not correspond to an azimuthally uniform source; rather the source strength follows the compact object and varies by something like a factor of five round the ring. Orbital motion and radial expansion are orthogonal, so the appearances of an orbiting and an expanding ring of fire differ in the phase of the pattern generated by the periodic variation in irradiation. For an orbiting ring, material moving blueward fastest is most strongly irradiated when the compact object is moving toward us at greatest projected velocity, orbital phase 0.25. For a ring expanding about the centre of mass of the binary, material moving blueward fastest is most strongly irradiated when it lies between us and the compact object, orbital phase 0.5 in the usual convention. Thus the whole pattern, otherwise identical, would be shifted by one quarter of a period relative to the model for an orbiting ring. The observations are in phase with the rotating ring model and 90$^{\circ}$ ahead of an expanding ring. An expanding source would exhibit  a phase different from an orbiting source, easily visible in a comparison  of model with data (Figs. 5-7, see also Bowler 2010a). Furthermore, for excitation by irradiation from the compact object, the details of the two horned structures cannot be explained for a radiating ring located much further out from the binary system than a radius of $2A$. An expanding ring would have to be excited by some quite different mechanism, which by some extraordinary coincidence yields effects exactly like those predicted by this circumbinary disk model.



\section{Conclusions}

  This analysis has greatly strengthened the case for a circumbinary disk generating emission spectra from material orbiting at approximately 250 km s$^{-1}$ at a radius  close to $2A$. At the same time, it has placed formidable obstacles in the way of supposed alternatives. This is important because if the source is a circumbinary disk the accreting compact object  must have a mass of somewhere in the region of 20 M$_{\odot}$ and so the compact object is probably a rather massive stellar black hole. The corollary is that the orbital speed of the companion in this binary must be in excess of about 130 km s$^{-1}$ (as reported by Cherepashchuk et al 2005), rather than the value extracted from absorption spectroscopy by Hillwig \& Gies (2008) and by Kubota et al (2010); 58 km s$^{-1}$.  An extended campaign of absorption spectroscopy might be able to provide evidence for or against the suggestion that the origin of those absorption lines could be in orbiting circumbinary material (Bowler 2010a, 2010c).
  
    Perhaps my most important conclusion is that the case for the circumbinary ring is so strong that both observers and theorists should take seriously the possibility that the compact object is massive and not merely nudging neutron star territory. Fresh evidence, either for or against, is more likely to emerge if it is looked for.

\begin{acknowledgements}
I thank an anonymous referee for a number of interesting and helpful comments.
This extension of the work presented in Bowler (2010a) was stimulated by a number of comments made by an anonymous referee for that paper and separately by a personal communication from D.R.Gies.
I acknowledge the crucial importance of the Blundell/Schmidtobreick  sequence of observations, made possible by the grant of Director's Discretionary Time on the 3.6-m New Technology Telescope.

\end{acknowledgements}

\end{document}